\documentclass[10pt, conference]{IEEEtran}
\IEEEoverridecommandlockouts
\usepackage{cite}
\usepackage{amsmath,amssymb,amsfonts}
\usepackage{algorithmic}
\usepackage{graphicx}
\usepackage{textcomp}
\usepackage{xcolor}

\usepackage{orcidlink}
\usepackage{caption}
\usepackage{subcaption}
\usepackage{braket}

\usepackage{ascmac} % for shadebox
\usepackage{amsthm}
\usepackage{booktabs}
\theoremstyle{plain}

\newtheorem*{thm*}{Theorem}
\theoremstyle{definition}

\usepackage[framemethod=default]{mdframed}
\newtheorem{dfn}{Definition}
\newmdenv[
  backgroundcolor=gray!10,
  linecolor=black,
  linewidth=0.4pt,
  skipabove=\baselineskip,
  skipbelow=\baselineskip
]{defbox}

\newcommand{\boxdef}[2]{%
\begin{defbox}
\begin{dfn}[#1]
#2
\end{dfn}
\end{defbox}
}

\def\BibTeX{{\rm B\kern-.05em{\sc i\kern-.025em b}\kern-.08em
    T\kern-.1667em\lower.7ex\hbox{E}\kern-.125emX}}
\begin{document}

\title{Anticipating decoder side-channel attacks in fault-tolerant quantum computers\\
\thanks{
% $^*$ Both authors contributed equally to this research.\\
SS is supported by the UK Engineering and Physical Sciences Research Council [Grant Number EP/S021582/1] and the National Research Foundation, Singapore through the National Quantum Office, hosted in A*STAR, under its Centre for Quantum Technologies Funding Initiative (S24Q2d0009).
SN acknowledges support from the JST Moonshot R\&D Program Grant No. JPMJMS256G and a JSPS Overseas Research Fellowship.
}
}

\author{
\IEEEauthorblockN{Shashvat Shukla \orcidlink{0000-0002-2164-744X}}
\IEEEauthorblockA{\textit{Department of Physics \& Astronomy} \\
\textit{University College London}\\
London, United Kingdom \\}
\and

\IEEEauthorblockN{Dan E. Browne \orcidlink{0000-0003-3001-158X}}
\IEEEauthorblockA{\textit{Department of Physics \& Astronomy} \\
\textit{University College London}\\
London, United Kingdom \\}

\and

\IEEEauthorblockN{Shin Nishio \orcidlink{0000-0003-2659-5930}}
\IEEEauthorblockA{\textit{Department of Physics \& Astronomy} \\
\textit{University College London}\\
London, United Kingdom and\\
\textit{Graduate School of Science and Technology}\\ 
\textit{Keio University}\\
Yokohama, Japan\\
}
}

\maketitle

\begin{abstract}
As quantum computing emerges as an applied technology, there is a growing need to protect quantum computers against information security attacks. This work identifies a new class of side-channel attacks against fault-tolerant quantum computers, in which the syndrome data that is sent to the decoder system is used to infer which computation (logical circuit) is taking place on the quantum computer. Our work introduces the concept of \textit{gate fingerprints}, which describes those patterns present in syndrome data that indicate which logical operation took place on the quantum computer. We show different effects by which logical operations produce gate fingerprints by focusing on Clifford+T computation in the surface code. We then explore how gate fingerprint information can be used to make inferences about the circuits or algorithms run on a quantum computer. Our findings indicate that decoder systems can be a vector for side-channel attacks and thus to prevent this, decoder systems should either be secured or built by a trusted party.
\end{abstract}

\begin{IEEEkeywords}
Fault-tolerant Quantum Computing, Side-channel Attack, Quantum Computing Security, Untrusted Decoder, Syndrome Information
\end{IEEEkeywords}

\section{Introduction}

Quantum computing is an emerging computing platform that can run new kinds of efficient algorithms and lower the memory usage and running time for solving various problems~\cite{shor1994algorithms, grover1996fast}. Many proposed quantum algorithms will require quantum computers with large numbers of qubits running many gate operations~\cite{gidney2025factor, dalzell2023quantum}. These requirements are beyond the capability of existing quantum computers, particularly because quantum computers are vulnerable to the effects of noise and errors. Extra qubits need to be introduced to provide redundancy against errors. Fault-tolerant quantum computing (FTQC)~\cite{shor1996fault, gottesman1998theory} studies such techniques for performing large-scale computations on noisy quantum systems. This involves realizing the quantum computation on \textit{logical} qubits implemented using many physical qubits according to a quantum error correcting code (QECC)~\cite{shor1995scheme}. Errors that occur during the computation can then be identified and corrected in real time.

The real-time detection and correction of errors in a fault-tolerant quantum computer is a computationally intensive process in itself \cite{Battistel_2023, barber2025real}. To identify where errors occur, sets of multi-qubit measurements, called stabilizer measurements, are performed periodically during a computation. These measurements produce many bits of data, collectively called the syndrome data. The syndrome data is then processed by means of a decoding algorithm that seeks to identify the errors most likely to have produced the observed syndrome. Appropriate corrections are then applied to negate the effect of the error. While small-scale demonstrations of FTQC can be done using CPUs to perform decoding \cite{moses2023race}, in larger systems, the larger amount of data and the low-latency required of decoders necessitate the use of GPUs~\cite{ferraz2025gpu}, FPGAs~\cite{das2022lilliput}, ASICs~\cite{das2022afs}, or a combination of them~\cite{barber2025real} to perform the decoding procedure. As such, decoding syndrome data in fault-tolerant quantum computing is emerging as an engineering challenge of its own, for which a purpose-built system needs to be designed. 

More broadly, emerging FTQC and quantum-HPC architectures increasingly treat decoding as part of a specialized classical control stack, alongside compilation, orchestration, and hardware-aware runtime services, rather than as a monolithic component of the quantum processor itself~\cite{mohseni2024build, 11249874, seelam2026reference}. Therefore, quantum computers may be assembled by buying a decoder system from a vendor and connecting the quantum computer control hardware to the decoder system. This separation of concerns will allow the control hardware experts and the FTQC experts to focus on what they do best, while remaining interoperable through an abstraction layer \cite{husseini2025qeci_aps}. 

Prior work in FTQC has assumed that the decoder is a trusted and reliable component. For example, techniques to improve the decoding success rate by giving the decoder information about the quantum computation have been studied~\cite{cain2024correlated, wu2024lego, zhou2025learning, Serra26decoding}. While it is likely that optimal decoder performance will require that a decoder be given full information about the quantum circuit being executed, one can imagine scenarios where passing this information to the decoder represents a security risk. It is natural, therefore, to consider the performance and security properties of \textit{circuit-agnostic decoding}, where as little information as possible about a circuit is provided to the decoder.

In such a setting,  the decoupling of the decoder system from the rest of the quantum computing stack raises the question of new information security vulnerabilities that may arise if the decoder system is malicious, faulty or compromised. 
In this work, we consider a confidentiality threat model in which the decoder is assumed to perform its decoding task correctly, but is not trusted with respect to the confidentiality of the computation. We call this the circuit-agnostic decoding model. Concretely, the decoder is allowed to observe the spacetime syndrome presented at the decoder interface and attempt to infer properties of the hidden logical computation from it, but it does not actively tamper with the decoding output, inject faults, or deliberately return incorrect corrections. This models both an honest-but-curious~\cite{paverd2014modelling} decoder vendor and a passively compromised decoder system.
To explore this risk, our work investigates how information about the logical circuit running on the quantum computer leaks to the decoder. Of course this will happen if information about the circuit is explicitly given to the decoder to improve its performance, so we focus on scenarios where one is trying to give the decoder as little information as possible to perform its function. We show in two natural settings that the decoder system can infer details about what computation is taking place in the quantum computer from just the syndrome information that it receives, even if it is not given explicit details of the computation. Our main finding is that although one may be inclined to think of syndrome data as data about background noise that needs to be corrected away, there are in fact \textit{gate fingerprints}\footnote{
The term fingerprint has also been used in prior quantum-computing security work, but with different referents and objectives. Mutolo et al. use error-syndrome statistics as a fingerprint of hardware/backend identity for authentication~\cite{mutolo2025quantum}, whereas MacNeil et al. use localized noise profiles as a fingerprint for circuit-tampering detection~\cite{macneil2025authenticating}. By contrast, in this paper, a gate fingerprint refers specifically to gate-dependent information that leaks to the decoder via syndrome data.
} that logical gates leave in the syndrome data from which an untrusted decoder system is able to gain information about the computation occurring in the quantum computer that it supports. 

The information security of quantum computers is an emerging topic of research \cite{szefer2025researchdirectionsquantumcomputer}. While quantum computing as an applied technology is still in an early stage of development, there are a number of reasons to start thinking about quantum computer security early. Firstly, quantum computers are powerful, expensive and strategic technologies which means adversaries will have stronger incentives to attack them for example to get free time on a quantum computer, to steal intellectual property on quantum computer designs, or to damage the quantum computer. Secondly, the quantum computing stack is complex and different from the classical computing stack, making the attack possibilities entirely new and uncharted. Our project is also motivated by the hope that lessons from the history of classical computer security will be learned and applied in quantum computing so that mistakes of the past will not be repeated.

Until recently, the focus in the field of quantum computer security has been on NISQ computers, however it is shifting towards FTQC \cite{trochatos2025exploration}. Topics in FTQC security have included information leakage attacks against control hardware to obtain logical qubit access patterns \cite{trochatos2025trace}, and physical attacks to disrupt the integrity of decoding and quantum error correcting codes \cite{lenssen2025fooling}. A taxonomy of FTQC security vulnerabilities has been proposed \cite{trochatos2025exploration} to categorise these and other previously discovered attack types.

\begin{figure}
    \centering
    \includegraphics[width=\linewidth]{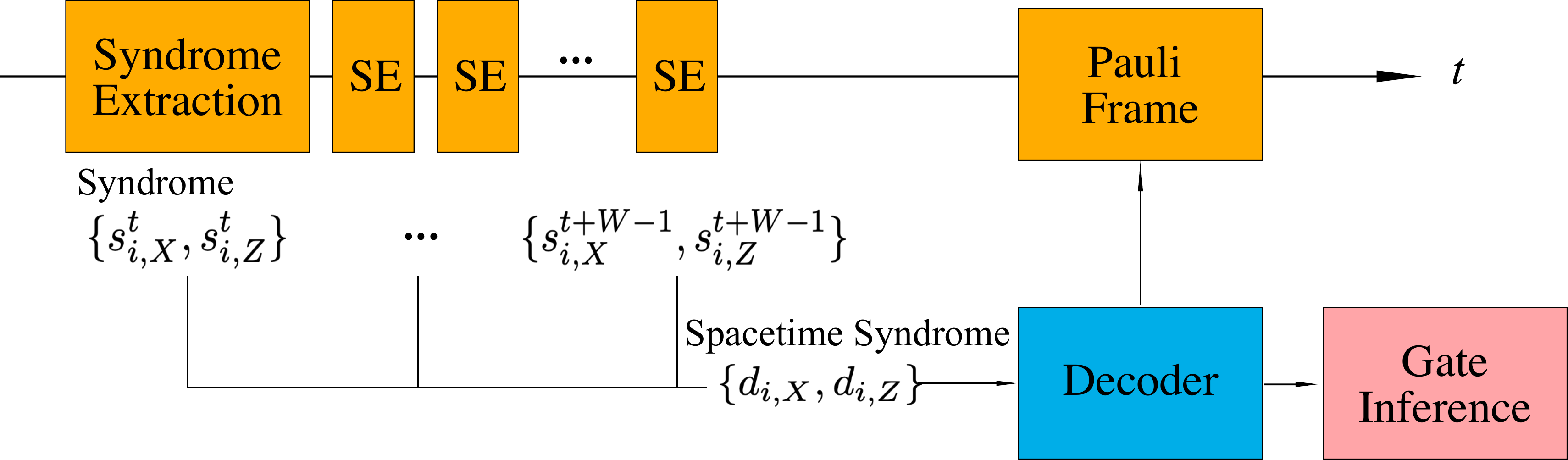}
    \caption{
    Threat model for decoder-side confidentiality leakage. The decoder is honest-but-curious, which is assumed to perform its task of estimating the Pauli frame from the spacetime syndrome, but is not trusted with respect to confidentiality. Even without being given an explicit circuit description, the information available at the decoder interface may be exploited to infer properties of the hidden logical computation, such as the logical gate applied in a given decoding window.
    }
    \label{fig:model}
\end{figure}

In this work, we identify a new type of side-channel attack in fault-tolerant quantum computers, where an untrusted decoder can exploit syndrome data to infer which logical operations are being executed. Fig.~\ref{fig:model} shows the threat model for the decoding system in a fault-tolerant quantum computer.

The remainder of this paper is organized as follows.
In Section \ref{sec:preliminaries} we present the formal model of syndrome data used in this work. 
In Section \ref{sec:gateinfer}, we formalize the information leakage vulnerability through the notion of \textit{gate fingerprints}, which are gate-implementation-dependent statistical patterns in spacetime syndrome data. 
We then analyze how such fingerprints arise in surface-code implementations of Clifford+$T$ computation, and show that they can reveal not only the type of logical gate, but in some cases also details of its physical realization. We also formulate the gate inference problem that an adversary performing this attack would have to solve, which is the task of estimating the logical gate from observed syndrome data. 
Section \ref{sec:circuitiden} describes techniques that allow the curious decoder to combine fingerprint information across the execution of a computation to determine the circuit or algorithm being run on the quantum computer. Section \ref{sec:conclusion} ends with security recommendations and ideas for future work.  

\newpage
\section{Preliminaries}
\label{sec:preliminaries}
In this section, we formalize the decoder-side leakage setting considered in this work. We assume that a logical quantum circuit is executed on a fault-tolerant quantum computer and that the decoder receives the spacetime syndrome required for real-time decoding. The decoder is assumed to perform its decoding task correctly, but it is not trusted with respect to confidentiality. It may attempt to infer properties of the hidden logical computation from the information available at the decoder interface. This captures both an honest-but-curious decoder supplied by an external vendor and a decoder system that has been passively compromised or is subject to information leakage.

Our primary threat model is passive. The decoder observes the spacetime syndrome and may additionally have access to decoder-side configuration information necessary for normal operation, but it does not inject faults, alter the reported decoding result, or deliberately return incorrect corrections. Thus, the attack surface studied here is a confidentiality side channel, not an integrity or availability attack on the decoding process.

We assume that the decoder is not given an explicit description of the logical circuit, the algorithm identity, or the intended gate sequence. However, this does not imply that the decoder interface is free of structural information. Even when explicit circuit metadata is withheld, the spacetime syndrome itself may implicitly reveal features such as patch extent, detector layout, or transient geometry changes arising from code deformation or lattice surgery. In a stronger variant of the model, the decoder may also be able to associate syndrome streams with logical qubit or code-block identities.

The adversary's primary goals are to infer properties of the logical gate sequence and to distinguish between candidate circuit families or algorithm classes. More detailed inferences, such as interaction structure or qubit mappings, are treated as downstream possibilities rather than primary claims of this paper. We further assume that the adversary may obtain reference fingerprints in advance, for example by profiling the same or a similar hardware stack, and may use these to aid gate inference.

\subsection{Syndrome extraction}
\boxdef{Syndrome}{
For each logical qubit $i$, let $\{s_{i,p,k}\}$ denote the set of $p$-type stabilizer generators with $p \in \{X,Z\}$ and index of stabilizers $k = 0,1,\dots,|G_p|-1$ where $|G_p|$ stands for the number of $p$-type stabilizers.

In the $t$-th round of syndrome extraction, the projective measurement of $s_{i,p,k}$ yields an eigenvalue
$m^t_{i,p,k} \in \{+1,-1\}$.

We define the corresponding binary syndrome bit
$$s^t_{i,p,k} \in \{0,1\}, \qquad s^t_{i,p,k} = \frac{1 - m^t_{i,p,k}}{2},$$
so that $s^t_{i,p,k} = 0 $ (resp. 1) indicates a $+1$ (resp. -1) measurement outcome.

The associated X-type syndrome vector at round $t$ for logical qubit $i$ is then
$$\mathbf{s}^t_{i,X}
= \bigl( s^t_{i,X,0}, s^t_{i,X,1}, \dots, s^t_{i,X,|G_X|-1} \bigr),$$
which is a binary vector whose length equals the number of X-type stabilizer generators. A Z-type syndrome vector $\mathbf{s}^t_{i,Z}$ is defined analogously.\\}

In the case where noise-free ideal syndrome measurements are possible, errors on qubits can be estimated and corrected based solely on the syndrome at each point in time. However, more realistically, measurements contain errors, so multiple rounds of syndrome measurements can be performed to estimate errors on the qubits and measurement errors \cite{dennis2002topological}. Such syndromes involving multiple rounds are called spacetime syndromes.

\boxdef{Spacetime syndrome}{
Fix a total number of syndrome-extraction rounds $T \ge 1$. For spacetime decoding, we define detection events, i.e., temporal changes in syndrome values.
We define the spacetime syndrome bits (detection events) as
$$d^{t+\frac12}{i,p,k}
:= s^{t+1}_{i,p,k} \oplus s^{t}_{i,p,k}$$ for $t = 0,1,\dots,T-1,$
where $\oplus$ denotes addition modulo 2.

The associated $p$-type spacetime syndrome vector is
$$\mathbf{d}_{i,p}
= \bigl( d^{t+\frac12}_{i,p,k} \bigr)_{t=0,\dots,T-1 \; k=0,\dots,|G_p|-1}
\in {0,1}^{T |G_p|}.$$

The set of all such vectors $\{\mathbf{d}_{i,p}\}_{i,p}$ is called the spacetime syndrome.
}

Here $d^{t+\frac12}_{i,p,k} = 1$ if and only if the outcome of stabilizer $s_{i,p,k}$ flips between round $t$ and $t+1$.

\boxdef{Detector check matrix~\cite{higgott2025sparse}}{
Consider a fault-tolerant syndrome-extraction circuit with \(N\) possible elementary error locations, indexed by
\(\ell=1,\dots,N\), and let \(D\) be the total number of detector outcomes in the spacetime syndrome.
The \emph{detector check matrix} is the binary matrix
\[
H \in \{0,1\}^{D\times N},
\]
whose \((j,\ell)\)-entry is defined by
\[
H_{j\ell}=1
\quad\Longleftrightarrow\quad
\text{an error at location }\ell\text{ flips detector }j,
\]
and \(H_{j\ell}=0\) otherwise.

Equivalently, if
\[
e=(e_1,\dots,e_N)\in\{0,1\}^N
\]
denotes the binary error pattern, where \(e_\ell=1\) indicates that the elementary error at location \(\ell\) occurred, then the resulting detector outcome vector
\[
d\in\{0,1\}^D
\]
satisfies
\[
d = He \pmod 2.
\]
Thus, each column of \(H\) specifies the set of detectors triggered by a single elementary fault, and the full detector outcome is obtained by summing these columns modulo \(2\).
}

The detector check matrix provides a compact description of how physical faults are mapped to detector outcomes in a given fault-tolerant implementation. In particular, it depends not only on the underlying code but also on the syndrome-extraction circuit and gate implementation used during the decoding window. Therefore, if different logical operations are realized through different fault-tolerant circuits, the corresponding detector mappings can differ, which in turn can induce gate-dependent statistical structure in the observed spacetime syndrome. This observation motivates the notion of gate fingerprints introduced in the next section.

\section{Gate Fingerprints and Gate Inference}
\label{sec:gateinfer}
\subsection{Gate Fingerprint Definition}
In our setting, the decoder does not observe an abstract logical circuit directly; it observes the spacetime syndrome, i.e., the pattern of detector events generated across repeated syndrome-extraction rounds. Because these detector events are induced through a gate-dependent fault-tolerant implementation and its associated detector mapping $H(G)$, their distribution can depend on the executed logical gate. We refer to this gate-dependent statistical structure in spacetime syndrome data as a \emph{gate fingerprint}.

\boxdef{Gate Fingerprint in Spacetime Syndrome}{
Let $D \in \{0,1\}^m$ denote the spacetime syndrome (detector outcomes) observed over a fixed decoding window, and let $G \in \mathcal{G}$ denote the hidden logical gate or short logical schedule executed in that window. We say that the decoder interface exhibits a gate fingerprint if the conditional distribution $P(D\mid G)$ depends on $G$.
\label{def:fingerprint}
}

The notion of a gate fingerprint is useful for separating a physical phenomenon from a specific attack algorithm. At this stage, we only claim that detector outcomes can depend on the executed logical operation through the underlying fault-tolerant implementation. In other words, the decoder receives information that may carry a computation-dependent fingerprint.

\subsection{Gate Fingerprint on surface codes}
In this subsection, we instantiate Definition~\ref{def:fingerprint} on fault-tolerant logical operations implemented on surface codes.
The detector observes a spacetime pattern of detector events generated by a gate-dependent syndrome-extraction schedule and its associated detector check matrix.
This can be used to infer the logical gate information. Therefore, even when the logical circuit is specified at a high level, the conditional distribution of detector outcomes can depend on the particular fault-tolerant realization of the gate.

More concretely, a gate fingerprint may arise from several distinct sources:
\begin{enumerate}
    \item which stabilizer family is initially ``flagged'' by state preparation, i.e. either all X or all Z stabilizers will be fixed deterministically at the beginning or end of the decoding window, while the other will be uniformly random. (e.g. Sec.~\ref{subsub:preparation}),
    \item which data and ancilla qubits participate in the implementation (e.g. Sec.~\ref{subsub:Pauli}),
    \item how faults propagate through the syndrome-extraction circuit (e.g. Sec.\ref{subsubsec:CX}), and
    \item  whether the logical operation changes the geometry or boundary conditions of the code, as in code deformation or lattice surgery~\cite{horsman2012surface} (e.g. Sec.\ref{subsubsec:CX}).
\end{enumerate}
These mechanisms induce gate-dependent statistical structure in the spacetime syndrome, which can leak information about the logical circuit to the decoder.

In the remainder of this subsection, we illustrate these categories using
representative logical operations from the instruction set
$\{$Initialization of $\ket{0}$ and $\ket{+}$, $Id, X, Y, Z, H, S, \textrm{Transversal}\, CX,$ Lattice Surgery, $T \}$,
and discuss which aspects of the spacetime syndrome make them distinguishable to the decoder.

\subsubsection{State preparation}
\label{subsub:preparation}
First, we consider cases in which the gate fingerprint is easily identifiable.
$\ket{0}_L$ state and $\ket{+}_L$ state are fundamental resources for quantum computing. 
For logical state preparation in a rotated surface code, the difference between $\ket{0}_L$ and $\ket{+}_L$ initialization comes from which type of stabilizer is fixed by the initial product state.

For the initialization of $\ket{0}_L$ state, we first initialize all the physical qubits as $\ket{0}^{\otimes N}.$ For any $Z$-type stabilizer
$S_Z=\prod_{i\in p} Z_i,$
we have
\[
S_Z \ket{0}^{\otimes N}=\ket{0}^{\otimes N},
\]
because $Z_i\ket{0}=\ket{0}$. Therefore, all $Z$-stabilizers yield the default outcome $+1$, without considering the effects of noise. By contrast, for any $X$-type stabilizer $S_X=\prod_{i\in p} X_i,$
$$
S_X \ket{0}^{\otimes N}\neq \pm \ket{0}^{\otimes N},
$$
so $\ket{0}^{\otimes N}$ is not an eigenstate of $S_X$. One can check that the first $X$-stabilizer measurements can return $\pm1$ uniformly at random.

Similarly, for the initialization of $\ket{+}_L$ state, all data qubits are initialized in $\ket{+}^{\otimes N}$,
then
$$
S_X \ket{+}^{\otimes N}=\ket{+}^{\otimes N},
$$
because $X_i\ket{+}=\ket{+}$. Thus all $X$-stabilizers yield $+1$. On the other hand,
$$
S_Z \ket{+}^{\otimes N}\neq \pm \ket{+}^{\otimes N},
$$
since $Z_i\ket{+}=\ket{-}$. Like before, the first $Z$-stabilizer measurements return $\pm1$ uniformly at random.

In summary,
$$
\text{On }\ket{0}^{\otimes N},
Z\text{-stabilizers give } +1,\quad X\text{-stabilizers give } \pm1,
$$
$$
\text{On }\ket{+}^{\otimes N},
X\text{-stabilizers give } +1,\quad Z\text{-stabilizers give } \pm1.
$$

Therefore, the initialization fingerprint is not merely a change in average detector rate, but a structural asymmetry in \emph{which stabilizer family is flagged at the boundary of the decoding window}, hence it is easy to distinguish between the two.

\subsubsection{Identity gate}
For the identity gate, the fingerprint is equivalent to the system's background noise. If the background noise has distinctive characteristics, it can sometimes make the differences clearer to work with by subtracting the fingerprint of the identity gate from any chunk of syndrome data before looking for gate fingerprints.

\subsubsection{Pauli $X,Y,Z$ gates}
\label{subsub:Pauli}
The logical Pauli $X$, $Y$, and $Z$ gates will produce fingerprints with more triggered detectors at the time when the gate is applied because of the physical gate errors from applying a physics $X$, $Y$, or $Z$ gate to each physical qubit. Note that the application of physical gates typically introduces more errors than doing nothing (i.e. identity gate). 

Since all the Pauli gates are applied to the same qubits, under the non-biased channel where $X$ errors occur with the same probability as $Z$ errors, it is impossible to distinguish these three gates.

\subsubsection{Hadamard and Phase gate}
The Hadamard and Phase gates have fingerprints arising from how they propagate $X$ and $Z$ errors in time. 

The logical Hadamard exchanges $X$-type and $Z$-type logical operators~\cite{chen2026transversal}. In the syndrome data, this can be seen as a conversion of past $X$ ($Z$) errors to $Z$ ($X$) errors. Consequently, a decoder may infer the presence of a Hadamard gate from an exchange in the syndrome statistics between the two detector families.

% H: X-> Z Z->X
% S: X-> X&Z Z-> Z
    % X is preserved
    % new Z is a combination
The phase gate $S$ transforms $X$ errors into $Y$ errors while leaving $Z$ errors unchanged. Therefore, one can find a time correlation between those checks as $Z$ check after the gate is now a combination of $X$ and $Z$ errors before the gate, while $X$ checks are preserved during the gate up to background noise.
Therefore, $S$ may remain distinguishable from both identity-like behavior and Hadamard-like behavior.

\subsubsection{$CX$ gates}
\label{subsubsec:CX}
Two-qubit logical gates can produce stronger fingerprints than single-qubit gates, because they create correlations across multiple logical patches. Here we compare two representative implementations of logical $\textit{CX}$, a transversal implementation and a lattice-surgery-based implementation.

\begin{figure}
    \centering
    \includegraphics[width=0.5\linewidth]{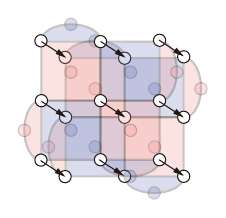}
    \caption{Transversal logical $\textit{CX}$ gate from foreground codeblock as a control qubit to background codeword as a target qubit. Each white vertex is a data qubit, and each red (blue) face is an $X$ stabilizer check ($Z$ stabilizer check). Each arrow illustrates the physical $\textit{CX}$ gate. $X$ errors propagate in the direction of the arrow, while $Z$ errors propagate in the opposite direction.}
    \label{fig:transversal}
\end{figure}

\begin{figure}
    \centering
    \includegraphics[width=\linewidth]{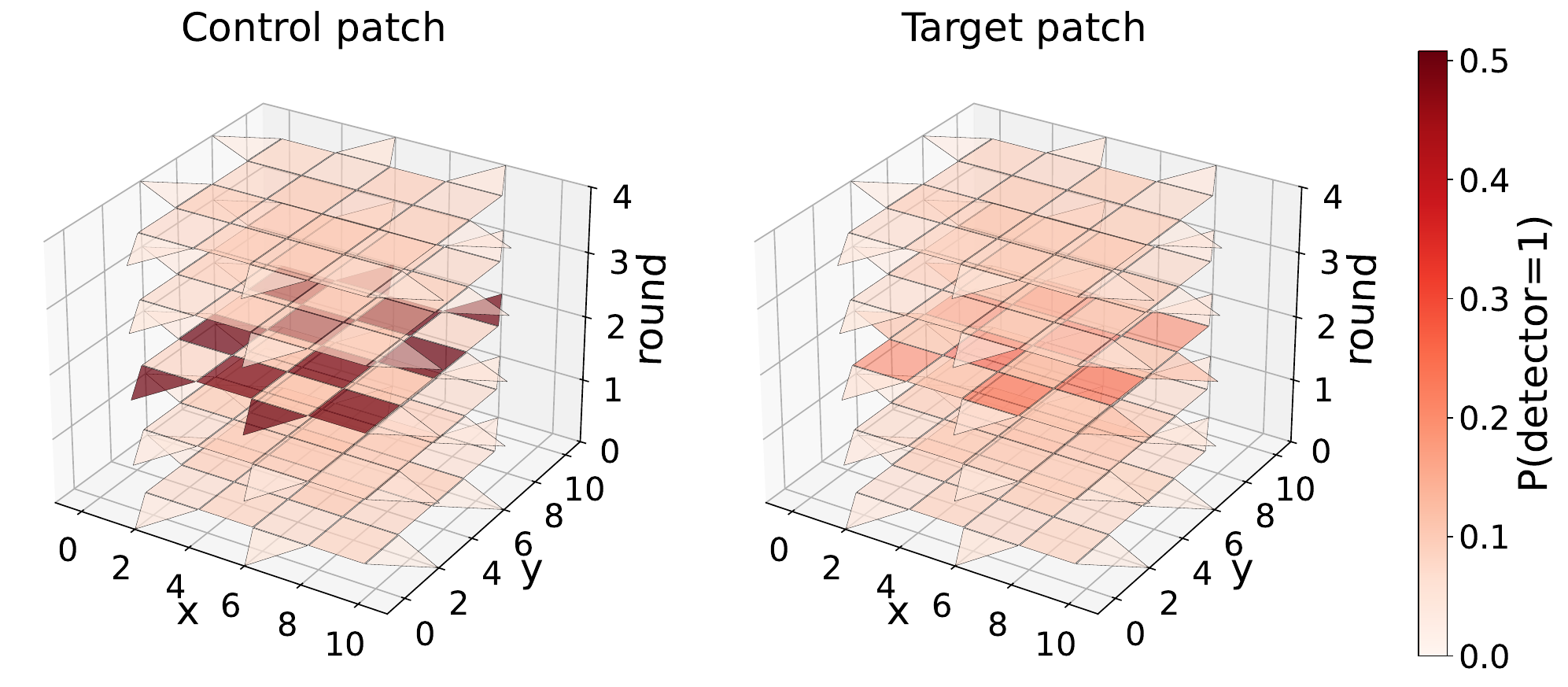}
    \caption{Gate fingerprint of transversal $CX$ gate on a pair of $d=5$ rotated surface code patch. Detector-event probabilities are shown over space and time for the control and target patches. The resulting signature is localized in time but spread over both patches, consistent with the coordinated two-block interaction of the transversal implementation. We set the physical error rates to \(p_{\mathrm{idle}}=10^{-3}\), \(p_{1q}=5 \times 10^{-3}\), \(p_{2q}=10^{-2}\), \(p_{\mathrm{reset}}=10^{-3}\), and \(p_{\mathrm{meas}}=10^{-3}\).}
    \label{fig:transversalcx_fingerprint}
\end{figure}

In a transversal logical $\textit{CX}$, corresponding physical qubits in the control and target code blocks interact pairwise, as illustrated in Fig.~\ref{fig:transversal}. This produces a characteristic pattern of cross-block participation during a localized time window. Although the overall code geometry is unchanged, the syndrome statistics can still reflect the presence of a coordinated two-block interaction. In particular, faults occurring during the transversal interaction can propagate along these pairwise couplings and induce correlated detector events across the two logical patches. The resulting fingerprint is therefore temporally localized, but spatially distributed over both code blocks. Fig.~\ref{fig:transversalcx_fingerprint} shows the gate fingerprint of the transversal $CX$ gate. Which of the $X$ and $Z$ stabilizers gets excited is not symmetric with respect to the direction of the $CX$ gate; the direction of the $\textit{CX}$ gate can also be inferred from the fingerprint.

\begin{figure}
    \centering
    \includegraphics[width=\linewidth]{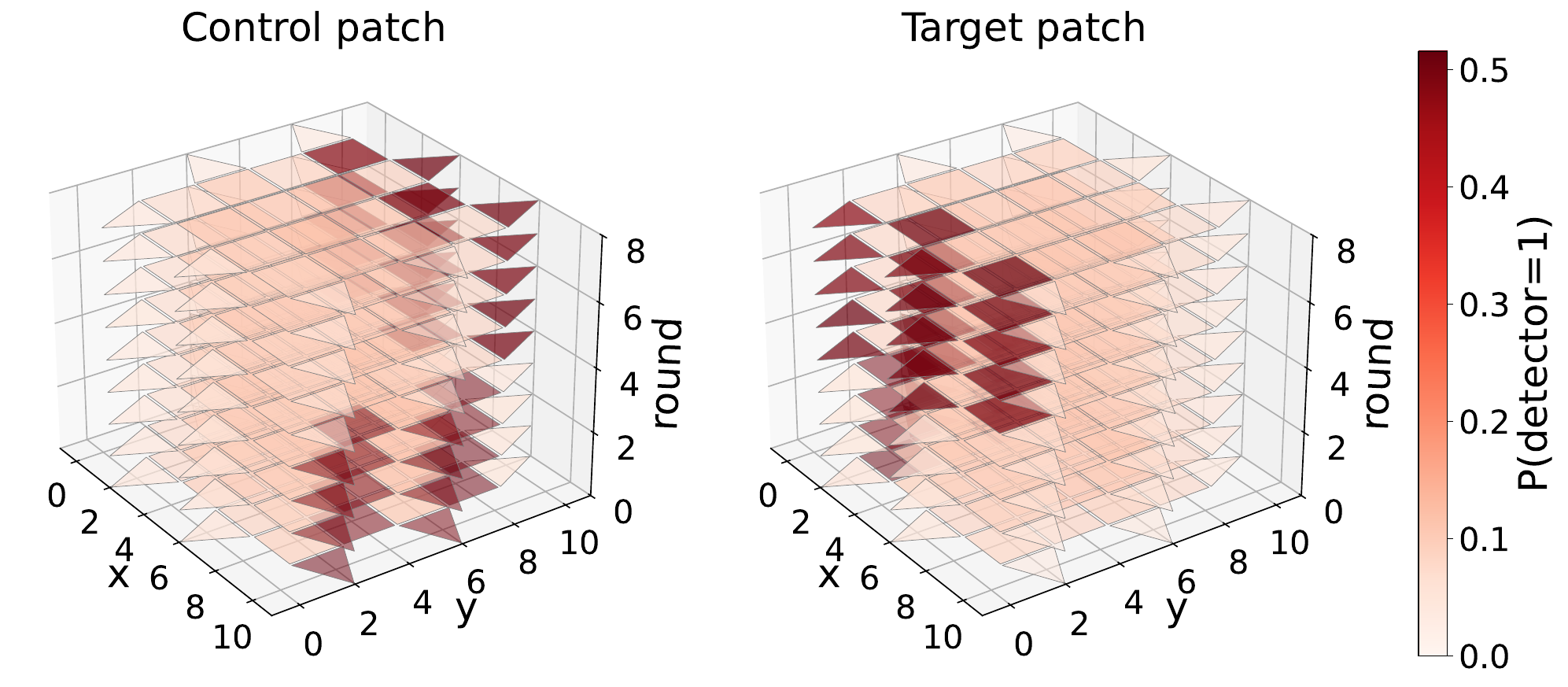}
    \caption{Gate fingerprint of lattice surgery-based $CX$ gate. The boundary associated with the $XX$ ($ZZ$) measurement is highlighted by a localized increase in detector-event probability. }
    \label{fig:ls_fingerprint}
\end{figure}

By contrast, in a lattice-surgery-based logical $\textit{CX}$, the gate is implemented through logical two-qubit Pauli measurement. In this case, the fingerprint does not arise merely from pairwise interactions between corresponding qubits. Instead, it arises from a temporary change in patch connectivity, boundary conditions, or effective code geometry. As a result, the decoder may observe detector patterns concentrated near the surgery region as shown in Fig.~\ref{fig:ls_fingerprint}, together with transient events associated with the merge and split stages. These features can be qualitatively different from those of the transversal implementation.

This comparison highlights an important point that the decoder may be able to distinguish not only between different logical gate labels, but also between different fault-tolerant realizations of the same logical gate. In other words, the object inferred by the decoder is not purely the logical semantics of $CX$, but the logical operation together with how it is physically implemented.

\subsubsection{$T$ gate}
The logical $T$ gate is a non-Clifford operation and, in surface-code-based FTQC, is typically implemented by magic-state injection\cite{bravyi2005universal, gottesman1999demonstrating, zhou2000methodology, litinski2019game}, with magic states obtained from magic state distillation~\cite{knill2004fault, bravyi2005universal} or cultivation~\cite{gidney2024magic} factories. Magic state injection can be realized by one and two-qubit Pauli measurements and classical controlled single-qubit gates. Therefore, for our purposes, we regard the $T$ gate as a subroutine rather than a primitive operation.

\subsection{Gate Inference}
Decoding is the problem of inferring the appropriate correction operations from syndrome data. On the other hand, an untrusted decoder trying to exploit the information leakage vulnerability to learn about the computation is also trying to solve the problem of inferring which logical gate operations were applied on the logical qubits from syndrome data. This is possible because different logical operations have different implementations, affecting how errors arise and propagate on physical qubits, which means that the logical operation leaves a signature in the syndrome data. We first present this problem in a Maximum Likelihood formulation. 

\boxdef{Maximum-Likelihood Gate Inference}{
\label{def:mlgi}
Let $\mathbf{d} = \{\mathbf{d}_{i,p}\}_{i,p}$ denote the observed spacetime
syndrome obtained from $T$ rounds of stabilizer measurements.
Let $\mathcal{G}$ be the set of supported logical gates, and let $G = \{G_{i,u}\}_{i,u}$ be a candidate logical-gate schedule, where each gate $G_{i,u} \in \mathcal{G}$ has a fixed duration in syndrome-extraction cycles (also known as code cycles).

Each schedule $G$ determines a physical syndrome-extraction circuit, including its set of physical error locations, the corresponding noise probabilities, and the decoding matrix $H(G)$ relating physical errors to the spacetime syndrome.

The marginal posterior probability of a schedule $G$ given the observed syndrome $\mathbf{d}$ is
\[
P(G \mid \mathbf{d})
\propto
P(G)
\sum_{\substack{e \in \{0,1\}^{N(G)} \\ H(G)e = \mathbf{d}}}
P(e \mid G),
\]
where $N(G)$ is the number of error locations for $G$,
$P(G)$ is a prior over gate schedules,
and $P(e \mid G)$ is the product of location-wise noise probabilities.

The maximum-likelihood estimate of the logical gate schedule is
\begin{equation} \label{eq1}
\begin{split}
\hat G 
& = \arg\max_G P(G \mid \mathbf{d})\\
& = \arg\max_G
\left[
P(G)
\sum_{\substack{e:\\ H(G)e = \mathbf{d}}}
P(e \mid G)
\right].
\end{split}
\end{equation}
}

Definition~\ref{def:mlgi} gives the most general form of the adversary’s task: inferring a hidden logical-gate schedule from the observed spacetime syndrome. In practice, however, solving Eq. (1) exactly is generally intractable for long schedules, since both the latent gate sequence and the associated physical fault configurations must be marginalized over.

We therefore begin with a restricted setting that captures the essential leakage phenomenon while being easy to simulate. Specifically, we consider a single decoding window containing one logical operation drawn from a finite gate set. In this case, gate inference reduces to a multiclass classification problem: given the observed spacetime syndrome $d$, estimate which logical gate most likely generated it. 

\begin{figure}
    \centering
    \includegraphics[width=0.7\linewidth]{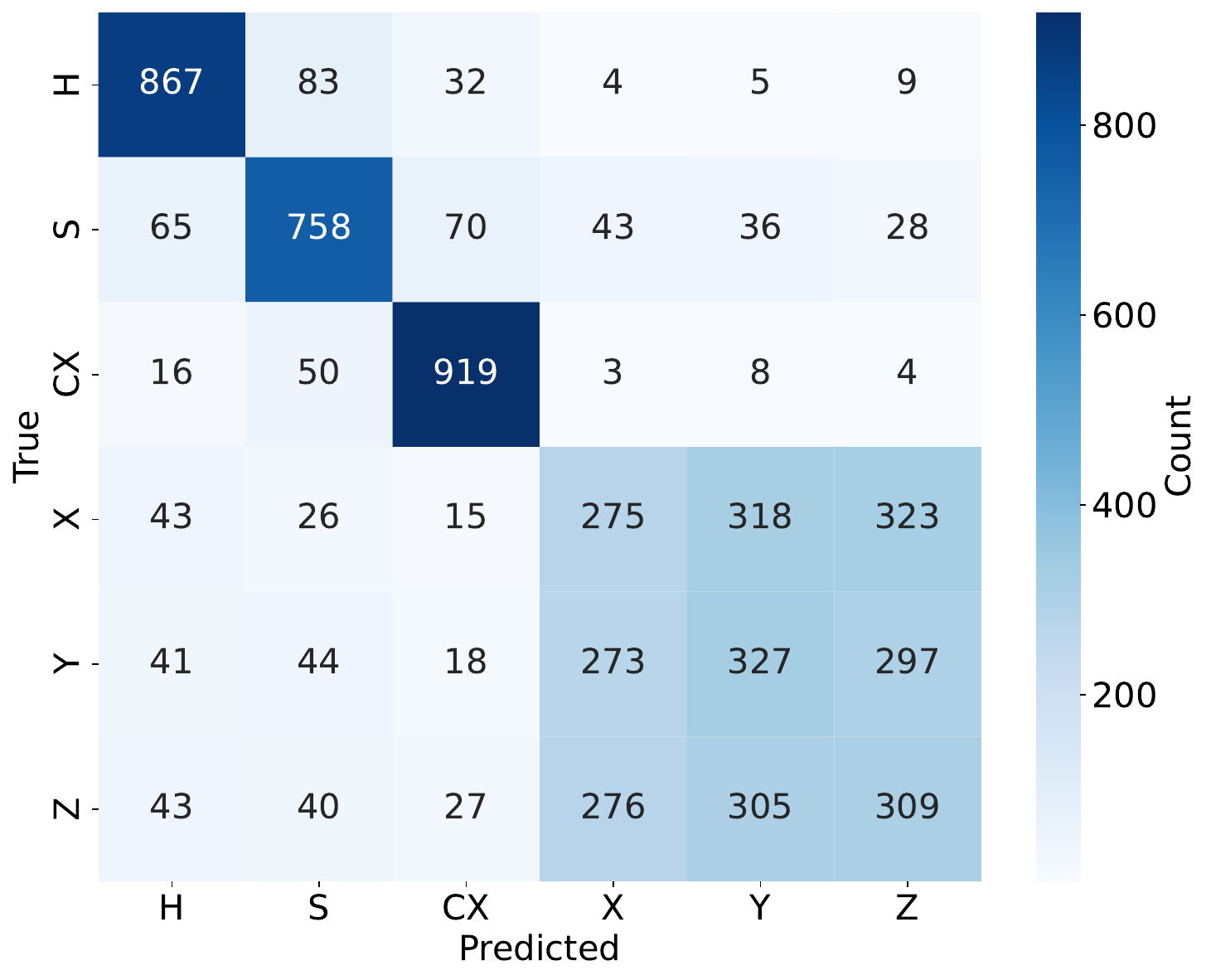}
    \caption{Classification metrics for logical gates. $H,S$, and lattice surgery exhibit distinctive fingerprints and are therefore generally classified correctly, whereas the Pauli gates $X$, $Y$, and $Z$ share the same fingerprint under the unbiased Pauli channel, rendering them indistinguishable. $1000$ shots per logical gate and $10000$ shots for training.}
    \label{fig:classification}
\end{figure}

Fig.~\ref{fig:classification} shows a representative confusion matrix for gate inference over the logical gate set.
For this simulation, we do not implement the exact maximum-likelihood inference of Eq.~(1). Instead, we use a naive-Bayes classifier on detector outcomes as a tractable approximation.
The classification task takes the gate fingerprints of two qubits as input. 
The results suggest that $H$, $S$, and $CX$ induce more distinctive fingerprints, whereas $X$, $Y$, and $Z$ are much harder to separate from one another under the unbiased Pauli noise model. Therefore the gate fingerprints are not uniformly distinguishable, but some logical operations can nevertheless be identified from syndrome data reliably. This is already sufficient to establish decoder-side information leakage. Moreover, an attacker does not need perfect single-gate recovery. Partial gate-level evidence can still be exploited when combined across time or matched against a restricted set of candidate circuits. 

Having established this point at the single-gate level, we now turn to how an adversary can aggregate such evidence across time to infer larger circuit structure in the next section.

\section{Circuit identification techniques}
\label{sec:circuitiden}

The previous section showed how syndrome information contains fingerprints of individual logical gate operations. This section explores the techniques that can be used to process the individual gate fingerprints to make inferences about the circuit or algorithm that is running in the quantum computer. As we have seen in the last section, gate inference is not deterministic and so we will not always be able to perfectly reconstruct the circuit that was run. However given that the set of known practical quantum algorithms is currently small \cite{dalzell2023quantum}, and that many quantum circuits have repeating structure, this gives us strong priors together with which information from gate fingerprints can be very revealing.

% We consider two scenarios, which correspond to different levels of information provided to the decoder: qubit-unaware decoding and qubit-aware decoding.
We now consider two variants of the decoder-side threat model introduced above, corresponding to different amounts of identifying information exposed at the decoder interface. In the weaker variant, which we call qubit-unaware decoding, the decoder receives syndrome instances without persistent logical-qubit or code-block identities. In the stronger variant, qubit-aware decoding, the decoder can associate syndrome streams with particular logical qubits or patches. The latter more closely reflects a straightforward implementation of decoder bookkeeping, while the former models an interface that attempts to reduce explicit metadata exposure. In both cases, however, the decoder may still exploit gate fingerprints present in the spacetime syndrome itself.

\subsection{Qubit-unaware decoding}

In this scenario, the quantum computer sends syndrome data to the decoder without associating the data to particular logical qubits. The decoder can still solve the decoding problems it is given and return the result to the quantum computer, however the problems are jumbled so that the decoder does not know which code block or qubits the data is associated with. This is a model of the minimal set of information that is given to a decoder without using sophisticated techniques to deliberately prevent decoder side-channel attacks. 

In this scenario, the decoder will still be able to observe gate fingerprints in the syndrome data it processes and estimate the count of each logical gate type. Certain subroutines such as adders \cite{cuccaro2004ripplecarry, takahashi2009quantum}
can have fixed ratios of gates across different problem sizes because they use a standard theoretical construction. Therefore, it is possible that for some circuits, a simple inference technique would be to match the ratios of gate counts observed in a circuit run to the gate statistics of a small set of candidate algorithms. However, this technique is likely to be of limited applicability as differences in compilation techniques can greatly change the ratio of gates. Many algorithms also do not preserve their gate ratios linearly in problem size. In particular, the fidelity of discretizations of rotation gates, by algorithms such as the Solovay-Kitaev algorithm \cite{dawson2005solovaykitaevalgorithm}, can totally change the ratio of gates. 

% \begin{table}[t]
% \centering
% \caption{Gate ratios for quantum subroutines. \red{Data is fake, needs to be updated! Maybe a stack based plot. Remove the distinction between theoretical and compiled. 15/04 }}
% \label{tab:gate-ratios}
% \begin{tabular}{lp{6.5cm}}
% \toprule
% Circuit / Algorithm & Gate Ratio \\
% \midrule
% \multicolumn{2}{l}{\textit{Theoretical construction}} \\
% Cuccaro et al Adder          & 7\,H\;:\;14\,T/T$^\dagger$\;:\;22\,CX \\
% Draper QFT Adder       & 24\,H\;:\;56\,T/T$^\dagger$\;:\;28\,CX\;:\;8\,S \\
% Square Root             & 12\,H\;:\;40\,T/T$^\dagger$\;:\;64\,CX\;:\;6\,X \\
% Quantum Multiplier     & 34\,H\;:\;192\,T/T$^\dagger$\;:\;228\,CX\;:\;8\,X \\
% \midrule
%     \multicolumn{2}{l}{\textit{Compiled benchmarks (MQTBench / FTCircuitBench)}} \\
% Grover's ($n=10$, MQTBench)          & 43\,H\;:\;128\,T/T$^\dagger$\;:\;97\,CX\;:\;15\,S \\
% Grover's ($n=10$, FTCircuitBench)    & 51\,H\;:\;204\,T/T$^\dagger$\;:\;83\,CX\;:\;22\,S \\
% QPE ($n=8$, MQTBench)                & 36\,H\;:\;88\,T/T$^\dagger$\;:\;72\,CX\;:\;12\,S\;:\;4\,X \\
% QPE ($n=8$, FTCircuitBench)          & 36\,H\;:\;152\,T/T$^\dagger$\;:\;64\,CX\;:\;8\,S \\
% QAOA MaxCut ($n=12$, MQTBench)       & 28\,H\;:\;168\,T/T$^\dagger$\;:\;54\,CX\;:\;6\,X \\
% VQE Ansatz ($n=8$, MQTBench)         & 16\,H\;:\;72\,T/T$^\dagger$\;:\;48\,CX\;:\;8\,R$_z$ \\
% Shor's ($n=9$, FTCircuitBench)       & 62\,H\;:\;310\,T/T$^\dagger$\;:\;194\,CX\;:\;18\,S\;:\;9\,X \\
% \bottomrule
% \end{tabular}
% \end{table}

More powerful inferences can be made by considering that decoders operate in real-time and thus the gate fingerprints in the syndrome data arrive in the same order as the circuit being run. Analysing gate statistics as they change can reveal algorithmic structure.

We performed a model simulation for three quantum circuits (Amplitude Amplification, HHL \cite{harrow2009quantum} and Quantum Fourier Transform). We implemented Amplitude Amplification, but the other circuits were obtained from FTCircuitBench \cite{harkness2026ftcircuitbench}. These circuits were then compiled to a Clifford+$T$ gate set \cite{ross2016optimalancillafreecliffordtapproximation}. We make the assumption that all $T$-states are teleported into the circuit. This means that the decoder only needs to detect fingerprints of Clifford gates in the syndrome data. We also assume that the syndrome from the magic factories is not being used by the decoder system in its attack.

The Clifford gates were counted in each layer of the circuit and the confusion matrix from Figure \ref{fig:classification} was applied to the data to simulate the statistics as observed by the untrusted decoder after gate inference. 

\begin{figure}[htbp]
    \centering
    \includegraphics[width=\linewidth]{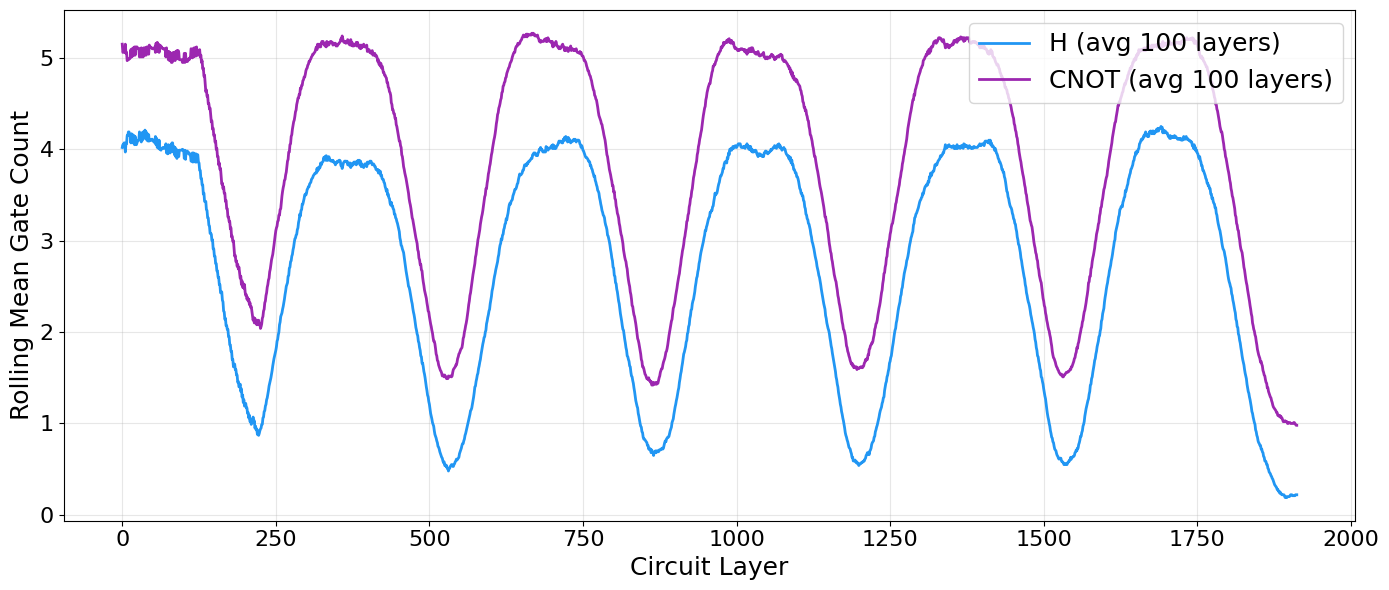}
    \caption{Rolling mean gate counts of the amplitude amplification circuit as seen by the decoder.}
    \label{fig:timebin_gatestats_aa}
\end{figure}

\begin{figure}[htbp]
    \centering
    \includegraphics[width=\linewidth]{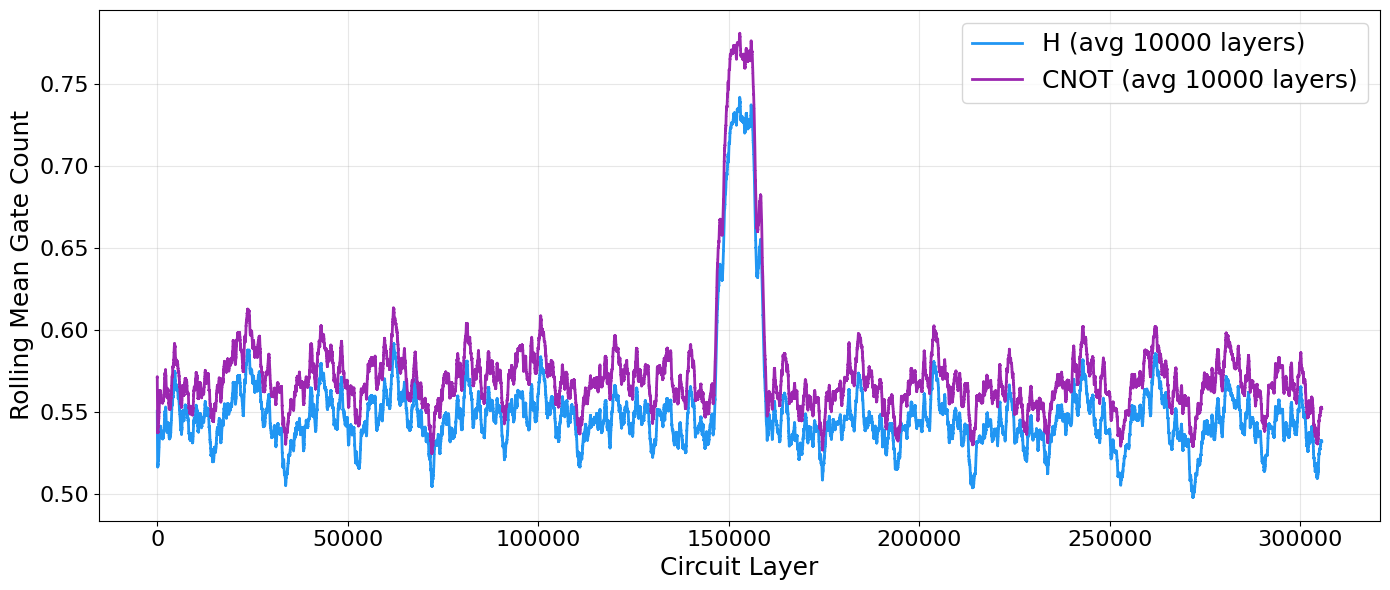}
    \caption{Rolling mean gate counts of the HHL algorithm as seen by the decoder.}
    \label{fig:timebin_gatestats_hhl}
\end{figure}

\begin{figure}[htbp]
    \centering
    \includegraphics[width=\linewidth]{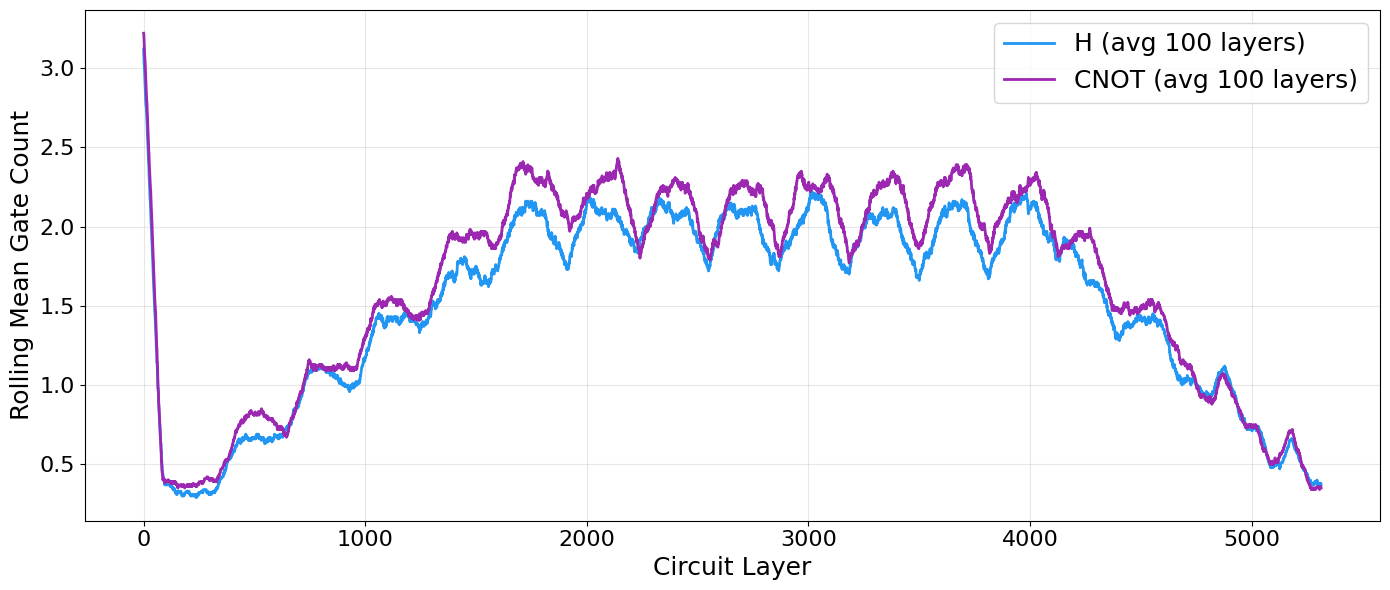}
    \caption{Rolling mean gate counts of the quantum Fourier transform circuit as seen by the decoder.}
    \label{fig:timebin_gatestats_qft}
\end{figure}

Figures \ref{fig:timebin_gatestats_aa}, \ref{fig:timebin_gatestats_hhl} and \ref{fig:timebin_gatestats_qft} show rolling average gate counts for just the $CX$ and Hadamard gates with simulated classification failures for Amplitude Amplification, HHL and the quantum Fourier transform, respectively. The two notable features of the plots are the changes in ratios of gates and the changes in the number of gates over time. The gate counts in Amplitude Amplification have a periodicity as it is an alternating operator algorithm. Such a structure would also be present in similar plots of Grover's algorithm and QAOA, as long as the two alternating operators have different gate ratios. In HHL, the plot captures the symmetry that arises from applying a unitary and its inverse, with gate-intensive arcsin arithmetic and controlled rotations in between. A notable feature of the quantum Fourier transform circuit is that the gate density starts low and ends low, but has higher gate density in the middle part of the circuit. This is explained by the order of gates in the QFT circuit. In the earlier part of the circuit, some qubits are waiting for operations on other qubits to complete before they are activated. Once all qubits are active, the circuit achieves maximal parallelisation and thus the rolling mean reaches a maximum. Then towards the end, the work on those qubits that started early is complete, and the parallelisation decreases as more qubits finish their work. 

Algorithms can be distinguished by inspection of the rolling mean gate counts across layers. To automate this inference a classifier could be built by handcrafted features such as periodicity and peaks, or sequence models could be explored.

There is even more information available to the decoder in the shape of the decoding problems. For example, performing lattice surgery between two patches further away creates a bigger surface code patch to be decoded. Further work can investigate attacks that use this information. 

\subsection{Qubit-aware decoding}

In this scenario, the decoder can associate syndrome data with specific logical qubits or code blocks. This is the default scenario that would arise in a straightforward implementation, where the decoder receives syndrome data labeled by logical qubit identity to facilitate bookkeeping of the Pauli frame. It may also have advantages such as tailoring the decoding to the precise noise models of the physical qubits involved in that logical qubit. While this simplifies implementation, it grants the adversarial decoder more information than in the qubit-unaware setting, enabling more precise circuit-level inference.

\subsubsection{Circuit interaction graph}

With access to most of the $CX$ gates that occurred in the circuit, and only a small number of false-positive $CX$ gates, an approximate structural analysis of the circuit will be possible to extract information such as the interaction graph of the circuit. The interaction graph is a weighted graph with qubits as vertices and higher-weight edges between qubits that have more two-qubit gates between them. This information can help classify the algorithm being used. For example, circuits corresponding to time evolutions of Hamiltonians have interaction graphs that reflect the local interactions within the Hamiltonian. Different algorithms and circuits are known to have different structures which have been studied for various reasons such as reconstructing circuit information from partial information obtained from side channels \cite{trochatos2025trace}, visualisation \cite{harkness2026ftcircuitbench} and investigating efficient circuit to device mappings \cite{bandic2025profiling}.

\subsubsection{Circuits with repeating structure}

\begin{figure}
    \centering
    \includegraphics[width=\linewidth]{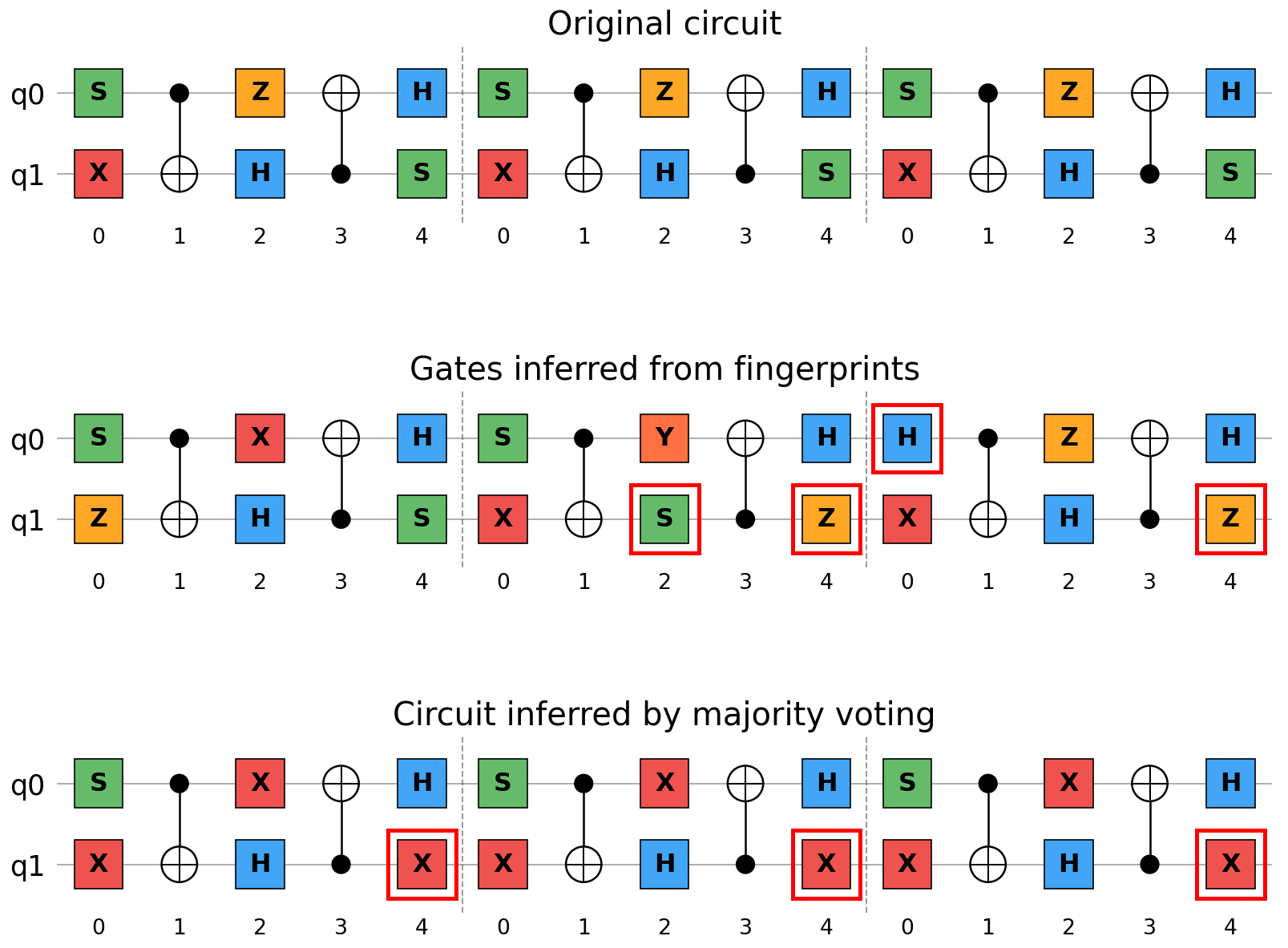}
    \caption{Imperfect gate inferences on a circuit with repeating structure can be reconstructed by majority voting. Note that due to the uniform confusion between X, Y and Z gates as in Fig. \ref{fig:classification}, the inferred circuit has all three replaced with the X gate.}
    \label{fig:repeating_illustration}
\end{figure}

\begin{figure}
    \centering
    \includegraphics[width=0.9\linewidth]{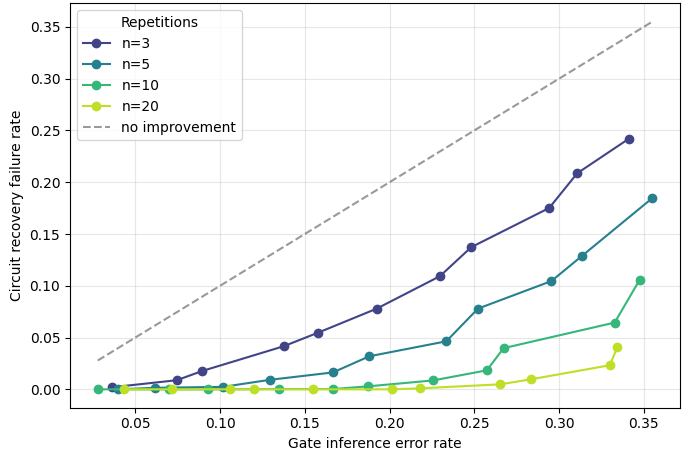}
    \caption{Performance of majority-voting technique for reconstructing repeating circuits after gate-wise fingerprint inference.}
    \label{fig:repeating_plot}
\end{figure}

Quantum Algorithms like Amplitude Amplification and Grover's algorithm with a repeating circuit structure are particularly easy to reconstruct from gate fingerprint information, because the repeating structure allows for the correction of any incorrectly inferred gates. Figure \ref{fig:repeating_illustration} illustrates this principle using a majority-voting approach for a simple circuit with a three-fold repeating structure. This approach is more successful when the circuit consists of a larger number of repetitions. Figure \ref{fig:repeating_plot} shows the performance of majority voting reconstruction as it depends on the number of repetitions in the circuit and the gate inference error. This technique can be particularly useful for identifying the oracle used in Grover's algorithm or Amplitude estimation, which reveals all the information about the problem being solved. 

\section{Conclusion}
\label{sec:conclusion}

In this work, we defined the circuit-agnostic model of quantum error correction decoding and studied its security properties. We identified a new class of side-channel attacks in fault-tolerant quantum computers, in which an untrusted decoder can exploit spacetime syndrome data to infer properties of the logical computation being executed. We formalized this leakage through the notion of gate fingerprints which are gate-dependent patterns that appear in the syndrome data as a result of distinctive noise processes of the underlying physical circuit implementation of the logical gate.

Focusing on Clifford$+T$ computation in the surface code, we showed that different logical operations can leave distinguishable signatures in the syndrome history. In particular, initialization procedures, Hadamard and phase gates, and multi-qubit operations such as transversal and lattice-surgery-based CX gates can produce characteristic patterns that may be exploited by a curious decoder to breach the confidentiality of the computation. Our analysis further suggests that the decoder may infer not only the logical gate label itself, but in some cases also features of its physical realization. These observations establish decoder-side inference as a concrete information-leakage threat in FTQC.

More broadly, our results show that syndrome data should not be treated as auxiliary information about environmental noise. Instead, it should be regarded as security-sensitive data which is circuit-dependent and whose exposure may reveal algorithmic or structural information about the user’s computation. 

This finding also reveals a tradeoff between information security and decoding performance. For the same reason that different logical operations leave fingerprints in their syndrome data, knowledge of the logical operation and how it is implemented allows the decoder to anticipate which patterns of errors are correlated, and how to interpret them. A number of works have thus explored how best to incorporate this knowledge into the decoder \cite{cain2024correlated, wu2024lego, zhou2025learning,Serra26decoding}. Designing the stack to prevent information leakage to the decoder thus means passing up on the advantages of circuit-aware decoding. 

Important directions for future work include quantifying the information content of gate fingerprints under realistic noise models and on other state-of-the-art FTQC architectures and compilation strategies such as generalized Pauli-based computing~\cite{litinski2019game} and high-rate Quantum LDPC code-based computing~\cite{yoder2025tour, baspin2025fast}. To check the details of the attack, a real-time demonstration should be implemented using real decoder systems and quantum computer hardware. Secure decoding architectures that balance fault-tolerance performance with confidentiality of computation also need to be explored and developed. In particular, this will require more techniques that can erase fingerprints from syndrome data. 

For further mitigation of the effect of the attack, verification methods in blind quantum computation~\cite{broadbent2009universal} can be useful. Also, it might be possible to hide the gate fingerprint by obfuscating the syndrome information for example using code automorphisms~\cite{koutsioumpas2025automorphism} or adding virtual errors. However, whether these strategies would be effective requires further research.

From a systems perspective, our findings imply that decoder systems should either be secured as trusted components or the interface to the decoder must be designed with explicit leakage-mitigation mechanisms. This is particularly important in cases where decoder systems are large high-performance computers themselves, which may mean they are vendor-supplied, shared, or connected to networks increasing the vulnerability to side-channel attacks. As it is early to know the precise details of such attacks, our only security recommendation at this stage is for users who do not want information about the quantum circuit they are running to be revealed to ensure that they trust the decoder system. 

\section*{Acknowledgement}
% Shin
SN thanks Yosuke Ueno for the useful discussion and for sharing references on fingerprint.
% Shashvat 
SS thanks SN and DB for teaching him fault-tolerant quantum computing. SS and SN thank George Umbrarescu for helpful discussions. 
% Dan

% \newpage
\bibliographystyle{IEEEtran}
\bibliography{main/main}
\end{document}